\documentclass[twocolumn,showpacs,preprintnumbers,amsmath,amssymb]{revtex4}

\usepackage{graphicx}
\usepackage{dcolumn}
\usepackage{bm}

\begin{document}

\preprint{}

\title{Transport properties of dense fluid argon}

\author{Sorin Bastea}
\email{bastea2@llnl.gov}
\affiliation{Lawrence Livermore National Laboratory, P.O. BOX 808, Livermore, CA 94550}

\date{18 September 2003}

\begin{abstract}
We calculate using molecular dynamics simulations the transport properties of realistically modeled fluid 
argon at pressures up to $\simeq 50GPa$ and temperatures up to $3000K$. 
In this context we provide a critique of some newer theoretical predictions for the diffusion coefficients of 
liquids and a discussion of the Enskog theory relevance under two different adaptations: modified Enskog 
theory (MET) and effective diameter Enskog theory. 
We also analyze a number of experimental data for the thermal conductivity of monoatomic and small 
diatomic dense fluids. 
\end{abstract}

\pacs{66.10.Cb, 66.20.+d, 66.60.+a}

\maketitle

Many real fluids are very well represented as ensembles of identical molecules interacting 
through pairwise spherical potentials and, except for more exotic versions of such interactions 
where molecular dynamics is still a very useful tool \cite{rsbs}, the thermodynamics of these systems 
has been well understood for a fairly long time in the context of various statistical 
mechanics theories \cite{tlwlw}. The transport properties on the other hand, i.e. self-diffusion coefficient, 
viscosity, thermal conductivity, are less amenable to accurate 
theoretical calculation and require computationally intense 
molecular dynamics simulations, hence the continuing interest in their study 
\cite{md96,re99,sh03}. Through a natural although {\it ad hoc} extension of the dilute gas Boltzmann equation, 
Enskog transport theory \cite{cc} provided the first prediction of the transport 
coefficients of the hard sphere fluid and opened the way to the calculation of transport properties of real 
dense fluids. Other, more heuristic theories have been also proposed, relying 
on general physical concepts such as``free-volume'' and ``caging'' \cite{ct,re02}, and excess-entropy scaling 
\cite{yr77,md96}. The predictive capabilities of all these methods are critically affected by both their intrinsec limitations 
and the additional interpretations required when they are applied to real fluids \cite{hmc,ks81}.
The molecular dynamics calculation of the transport coefficients of the hard sphere fluid model \cite{agw} 
has provided important insights on the limitations of the Enskog theory in the high density regime, 
as well as the connection between microscopics and hydrodynamics.
Experimental results on the transport properties of liquids at high pressures, e.g. tens of GPa, 
are only now becoming available for both molecular fluids \cite{asb} and liquid metals \cite{dd02}. 
Here we present molecular dynamics calculations of the transport properties of realistically 
modeled argon at pressures up to $\simeq 50GPa$ and temperatures up to $3000K$. 
In this context we provide a critique of some newer theoretical predictions 
for the diffusion coefficients of liquids and a discussion of the Enskog theory relevance under 
two different adaptations: modified Enskog theory (MET) and effective diameter Enskog theory. 
We also analyze a number of experimental data for the thermal conductivity of monoatomic and small 
diatomic dense fluids. 

Argon is generally believed to behave as a quintessential classical fluid in a rather wide 
range of densities and temperatures and it has been often modeled as a Lennard-Jones system \cite{vhc}. 
More accurate representations of its interactions are also available, e.g. the Barker-Fisher-Watts (BFW) potential 
\cite{bfw}, but they are more complicated and have been tested only at low pressures. A relatively 
simple pair-interaction that can account very well for the high density, high temperature thermodynamics 
of argon \cite{rmbx} is the Buckingham {\it exponential-6} potential: 
\begin{eqnarray}
u(r) = \epsilon\left[Ae^{-\alpha\frac{r}{r_{0}}}-B\left(\frac{r_{0}}{r}\right)^6\right]
\label{eq:exp6}
\end{eqnarray}
with well-depth $\epsilon$ corresponding to distance $r_{0}$, and $\alpha$ a numerical constant: 
$A=6e^{\alpha}/(\alpha-6)$, $B=\alpha/(\alpha-6)$, the general properties of which have been 
well studied \cite{ep98}. In addition 
to argon the {\it exponential-6} parametrization was shown to yield appropriate thermodynamics for other 
molecular fluids as well \cite{rr80}, e.g. $N_2$, $O_2$, $CO_2$, $CH_4$, $CO$ etc., particularly in the dense, 
hot regimes corresponding to shock-waves, detonations or planetary modeling \cite{hbl}. The understanding 
of such dynamic processes requires a knowledge of both the thermodynamic and transport properties of 
these fluids and their mixtures, 
and we study argon as a representative example. Other molecular fluids and the limitations of 
{\it exponential-6} modeling for small diatomics are also discussed in connection with 
available high pressure thermal transport experimental results.

We set $\epsilon/k_B=122K$, $r_0=3.85$\AA, $\alpha=13.2$ corresponding to argon \cite{rmbx}, 
and perform microcanonical 
(NVE) molecular dynamics simulations with $500$ particles (and some with $864$ particles) in a cube $L\times L\times L$
with periodic boundary conditions. The temperatures studied are $T=298, 1000, 3000 K$ and densities from slightly 
above the critical density ($\rho_c\simeq 0.54 g/cm^3$) to just below freezing, i.e. approximately $1.95 g/cm^3$ 
at $300$K, $2.75 g/cm^3$ at $1000K$ and $4.05 g/cm^3$ at $3000K$ (for comparison the triple point density is 
$\simeq 1.14g/cm^3$). The corresponding pressures are up to about 
$1.3GPa$, $9.3GPa$ and $52GPa$, respectively. The transport coefficients $D$ - (self-)diffusion coefficient, 
$\eta$ - shear viscosity and $\lambda$ - thermal conductivity, are calculated using 
the Green-Kubo formalism \cite{hm}. This entails determining the long time behavior of 
time integrals of auto-correlation functions of appropriate microscopic currents:
\begin{subequations}
\begin{eqnarray}
&&D=\lim_{t \rightarrow \infty}D(t)\\
&&D(t)=\int_{0}^{t}\langle v_{ix}(0)v_{ix}(\tau)\rangle d\tau\nonumber\\
&&\eta=\lim_{t \rightarrow \infty}\eta(t)\\
&&\eta(t)=\frac{1}{Vk_BT}\int_{0}^{t}\langle \sigma_{xy}(0)\sigma_{xy}(\tau)\rangle d\tau\nonumber\\
&&\lambda=\lim_{t \rightarrow \infty}\lambda(t)\\
&&\lambda(t)=\frac{1}{Vk_BT^2}\int_{0}^{t}\langle J^e_{x}(0)J^e_{x}(\tau)\rangle d\tau\nonumber
\end{eqnarray}
\end{subequations}
where $\hat{\mathbf{\sigma}}$ and $\mathbf{J}^e$ are the microscopic stress tensor and energy 
current, respectively, easily calculated in the course of molecular dynamics simulations:
\begin{subequations}
\begin{eqnarray}
&&\sigma_{xy}(\tau)=\sum_i \left[m_i v_{ix}(\tau)v_{iy}(\tau)+y_i(\tau)F_{ix}(\tau)\right ]\\
&&J^e_x(\tau)=\sum_i v_{ix}(\tau)\left\{\frac{1}{2}m_i v^2_i(\tau)+\frac{1}{2}\sum_{j\neq i}V_{ij}\left
[r_{ij}(\tau)\right]\right\}\nonumber\\
&&+\frac{1}{2}\sum_i\sum_{j\neq i}\left[x_i(\tau)-x_j(\tau)\right ]\mathbf{v}_i(\tau)\cdot \mathbf{F}_{ij}(\tau)
\end{eqnarray}
\end{subequations}

The two main sources of errors affecting the molecular dynamics calculation of transport coefficients are 
the system size - $L$ - and the time limit - $t_{lim}$ - on the calculation of the above time integrals.
Finite size effects on the calculation of transport properties have been extensively analyzed and indicate that systems 
of 500 particles typically yield results within $\simeq 2-3\%$ of the infinite system size extrapolations 
\cite{aw70,ew91,sh03}. The effect of an integration time limit $t_{lim}$ is more subtle and it has to do with the slow, 
algebraic decay 
of the Green-Kubo integrands \cite{aw70}. For usual, 3-dimensional systems these integrands behave at long times as 
$\rho_{\delta}(t)\propto k_{\delta}/t^{\frac{3}{2}}$ \cite{ehl,yp72}; $\delta$ stands for $D$, $\eta$ and $\lambda$. 
The factors $k_{\delta}$ have been calculated using the mode-coupling formalism and depend on the system
thermodynamics and on the transport coefficients themselves \cite{ehl,yp72}. We set $t_{lim}=0.95t_c$, 
where $t_c$ is the time needed for a sound wave to traverse the system, $t_c=L/c$, 
$c=(\partial p/\partial \rho)^{\frac{1}{2}}_s$ (adiabatic sound speed), and add the long time contributions to 
the final values of the transport coefficients. We find that these corrections can be as high as $12\%$ for the 
diffusion coefficient, in agreement with \cite{ew91}, up to $3\%$ for the viscosity, and smaller than $1\%$ for 
the thermal conductivity. For each thermodynamic point we run the simulations for $5-25$ million time steps, which 
corresponds, depending on density and temperature, to $3000-15000$ samples in the averaging of the 
auto-correlation functions.

As mentioned, we would like to compare the simulation results with available theoretical estimates. Among them 
the Enskog theory \cite{cc} is perhaps the best known; its predictions for the transport coefficients of the 
hard sphere system are:
\begin{subequations}
\label{eq:enskog}
\begin{eqnarray}
&&\frac{D_E}{D_B}=\frac{\rho b_{hs}}{y_{hs}}\\
&&\frac{\eta_E}{\eta_B}=\rho b_{hs}\left(\frac{1}{y_{hs}} + \frac{4}{5} + 0.7614y_{hs}\right)\\
&&\frac{\lambda_E}{\lambda_B}=\rho b_{hs}\left(\frac{1}{y_{hs}}+\frac{6}{5}+0.7574y_{hs}\right)
\end{eqnarray}
\end{subequations}
where $b_{hs}=2\pi\sigma^3/3$, $y_{hs}=p/\rho k_BT - 1$ and the pressure $p$ can be accurately calculated using the 
Carnahan-Starling equation \cite{hm}: $p/\rho k_B T=(1+\phi+\phi^2-\phi^3)/(1-\phi)^3$. 
The right-hand-side of the above equations depends only on the hard sphere 
packing fraction $\phi=\pi\rho\sigma^3/6$, while the left-hand-side contains the Boltzmann transport coefficients 
$D_B$, $\eta_B$ and $\lambda_B$, obtained in the limit of low densities \cite{cc}:
\begin{subequations}
\begin{eqnarray}
&&D_B=1.019\frac{3}{8\rho\sigma^2}\left(\frac{k_B T}{\pi m}\right)^{\frac{1}{2}}\\
&&\eta_B=1.016\frac{5}{16\sigma^2}\left(\frac{m k_B T}{\pi}\right)^{\frac{1}{2}}\\
&&\lambda_B=1.025\frac{75}{64\sigma^2}\left(\frac{k_B^3 T}{\pi m}\right)^{\frac{1}{2}} 
\end{eqnarray}
\end{subequations}
with $\rho=N/V$ the number density, $\sigma$ the hard sphere diameter and $m$ the molecular mass. 

The application of these results to real dense fluids requires suitable interpretation, and 
the so-called modified Enskog theory (MET) \cite{hmc} has been widely used.
The MET ingredients as applied to real fluids are i) the replacement of $y_{hs}$ with the ``thermal pressure'' 
of the fluid in question, $y=(\partial p/\partial T)_{\rho}/\rho k_B - 1$, which is then required to 
equal that of the hard sphere fluid, $y=y_{hs}$, and therefore leads by invoking the low density 
limit to ii) the identification of $b_{hs}$ with the second virial coefficient of the real fluid 
and its temperature derivative, $d[Tb(T)]/dT$, 
and iii) the replacement of $D_B$, $\eta_B$ and $\lambda_B$ with the real dilute gas transport coefficients of the fluid 
considered, $D_0$, $\eta_0$ and $\lambda_0$. The comparison of the MET predictions with experimental results 
for a variety of fluids up to densities about twice the critical density is rather favorable \cite{hmc}. 
However, a number of MET inconsistencies have been pointed out \cite{ks81}, and it is not clear if this 
approach would continue to be useful as the density is increased. For our MET estimates we recall that the 
transport coefficients of a dilute gas of molecules interacting through some general potential such as that 
defined in Eq. \ref{eq:exp6} 
can be written in the first Enskog approximation as \cite{cc}:
\begin{subequations}
\begin{eqnarray}
&&D_0=\frac{3}{8\rho r_0^2}\left(\frac{k_B T}{\pi m}\right)^{\frac{1}{2}}\frac{1}{\Omega^{(1,1)*}(T^*)}\\
&&\eta_0=\frac{5}{16r_0^2}\left(\frac{m k_B T}{\pi}\right)^{\frac{1}{2}}\frac{1}{\Omega^{(2,2)*}(T^*)}\\
&&\lambda_0=\frac{75}{64 r_0^2}\left(\frac{k_B^3 T}{\pi m}\right)^{\frac{1}{2}}\frac{1}{\Omega^{(2,2)*}(T^*)} 
\end{eqnarray}
\end{subequations}
where $\Omega^{(m,n)*}(T^*)$, $m,n=1,2$ are dimensionless collision integrals \cite{fk} depending on the interaction 
potential and reduced temperature $T^*=k_BT/\epsilon$, that we evaluate numerically. 
 
A different interpretation of the Enskog theory for real fluids has been advocated in \cite{ks81}, 
based on the use of a state-dependent hard sphere diameter directly in the Enskog relations Eqs. 
\ref{eq:enskog}. The success of statistical mechanics theories in predicting the thermodynamics 
of simple fluids \cite{bh76} is largely due to the idea of equivalent hard sphere diameters, which embody 
the dominant effect of short range repulsions on the structure and dynamics of liquids, particularly 
at high densities. The appeal of using the same diameter to calculate both the thermodynamic 
and transport properties of fluids lies therefore both in its simplicity and physical consistency. 
In the present work we adopt the definition of effective diameter 
provided by the Mansoori-Canfield variational method \cite{bh76}, which uses the fact that the 
first order perturbation theory approximation for the free energy of a system interacting through potential 
$u(r)$ is an upper bound for the free energy of the system. Using the hard sphere fluid as a reference 
this translates into:
\begin{eqnarray}
f_u(\rho, T) \leq f_{hs}(\phi, T) + 12\phi\int_1^\infty s^2g_{hs}(s; \phi)u(\sigma s)ds
\label{eq:vart}
\end{eqnarray}
where $g_{hs}(s; \phi)$, $s=r/\sigma$, is the pair-correlation function of the hard sphere fluid.
The optimal approximation for the free energy per particle $f_u$ (the interaction potential dependence 
is explicitly indicated for clarity) is obtained by minimizing the right hand side of Eq. \ref{eq:vart} with 
respect to $\sigma$, which provides at fixed density and temperature an effective hard sphere 
diameter. Thermodynamics is then derived in the usual way 
by taking the appropriate derivatives. A straightforward modification of the variational procedure Eq. \ref{eq:vart} 
\cite{mr79} further improves its accuracy for dense fluids \cite{tlwlw} and we use it for the 
calculation of the ``thermal pressure'' necessary for the MET estimates.

It is worth noting that the effective hard sphere diameter approach is not 
necessarily tied to the use of the Enskog theory, and can be interpreted more generally as a test of single-variable 
scaling for the transport properties of fluids modeled by realistic 
pair interactions. Similar ideas have been considered for the particular case of inverse 
power-law potentials \cite{ah75}. Perhaps even more importantly, this can also be viewed in the larger context of trying 
to uncover universal features of the transport properties of real fluids by mapping them into 
those of a reference system, e.g. hard sphere fluid. Using a so-called ``entropy packing fraction'' 
suggested by the variational method Eq. \ref{eq:vart}, a connection between transport 
coefficients and thermodynamic properties, specifically excess entropy (with respect to the ideal gas) 
per particle $s_{e}$, 
\begin{figure}
\includegraphics[width=8.0cm]{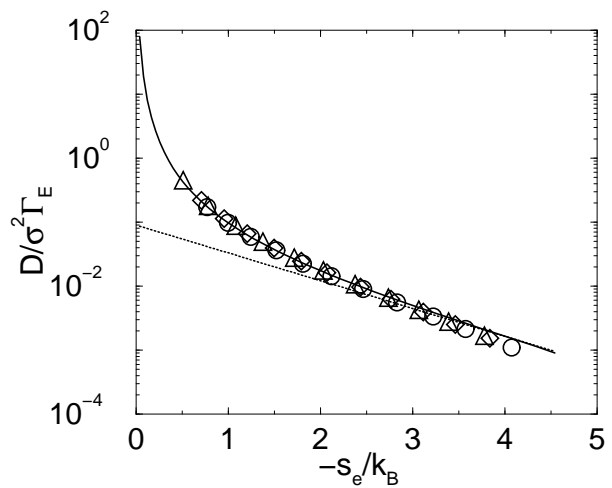}
\caption{Scaled diffusion coefficient of argon as a function of excess entropy (see text - Eq. \ref{eq:dif1}). 
Symbols: simulation 
results (circles - $T=298K$, diamonds - $T=1000K$, triangles - $T=3000K$). Solid line: hard sphere fluid 
diffusion coefficient from Ref. \cite{ew91}. Dotted line: Eq. \ref{eq:dif1} with $C=0.09$.}
\end{figure}
has been proposed more than 20 years ago \cite{yr77}. 
Another version of the excess entropy idea for the particular case of diffusion has been more 
recently suggested \cite{md96}, and we would like to analyze it here for the case of dense argon.  

The relationship between the diffusion coefficient and excess entropy postulated in \cite{md96} is:
\begin{eqnarray}
D^*=\frac{D}{\sigma^2 \Gamma_E}=C\exp{(s_{e}/k_B)}
\label{eq:dif1}
\end{eqnarray}
where the hard sphere diameter $\sigma$ and the Enskog collision frequency 
$\Gamma_E=4\sigma^2 g(\sigma)\rho\sqrt{\pi k_B T/m}$ are assumed to be the relevant length and time scale respectively, 
and $C$ is believed to be an universal constant. A certain definition for $\sigma$ and the contact value 
of the pair correlation function $g(\sigma)$, along with an approximation for $s_{e}$ has also been suggested for 
real systems \cite{md96}. Here we use the definition of $\sigma$ provided by Eq. \ref{eq:vart} and the values for 
$s_{e}$ and $g(\sigma)$ (which we denote by $g_c(\phi)$ to make explicit the dependence on $\phi$) obtained from the 
Carnahan-Starling equation of state \cite{hm}:
\begin{subequations}
\begin{eqnarray}
&&\frac{s_{e}}{k_B}=-\frac{4\phi-3\phi^2}{(1-\phi)^2}\\
&&g_c(\phi)=\frac{2-\phi}{2(1-\phi)^3}
\end{eqnarray}
\end{subequations}
Using this scaling the argon simulation results at the three temperatures studied are presented in Fig. 1 together 
with the diffusion coefficient of the hard sphere system \cite{ew91}; the universal curve proposed in \cite{md96}, 
i.e. Eq. \ref{eq:dif1}, is also shown. 

We find that single-variable scaling based on the effective hard sphere diameter of Eq. \ref{eq:vart} 
holds rather well, and the agreement with he hard sphere fluid diffusion coefficient is also reasonable. However, 
Eq. \ref{eq:dif1} appears to be valid only in a limited range of excess entropies $s_{e}$, as already remarked in 
\cite{cr00,jb02}. 
\begin{figure}
\includegraphics[width=8.0cm]{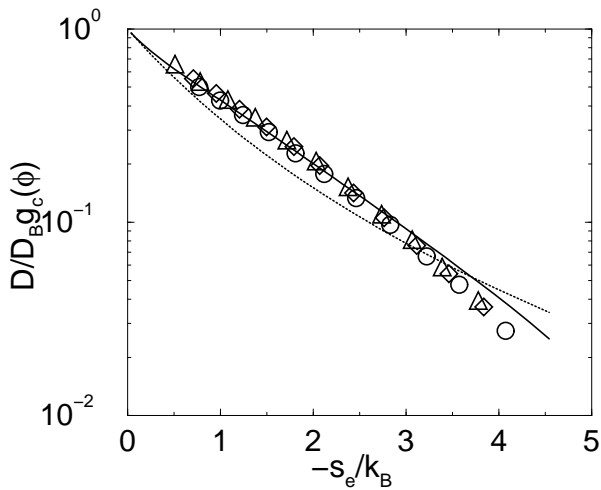}
\caption{Scaled diffusion coefficient of argon as a function of excess entropy (see text - Eq. \ref{eq:dif2}). 
Symbols (same as Fig. 1): simulation results. Solid line: hard sphere fluid diffusion coefficient from 
Ref. \cite{ew91}. Dotted line: (effective diameter) 
Enskog theory.}
\end{figure}
The discrepancy at lower (in absolute value) $s_{e}$, i.e. smaller packing fractions 
$\phi$, is severe and somewhat troublesome 
given that the Enskog theory, upon which the proposed relationship is loosely based, performs well precisely 
in that domain. To understand the problem with the scaling introduced in Eq. \ref{eq:dif1} we note that the left-hand-side 
of that equation can be written up to a multiplicative constant as $D/D_B g_c(\phi)\phi^2$. Therefore, in the limit of a dilute 
system, $\phi\rightarrow 0$, this term will diverge as $1/\phi^2$, while the right-hand-side of Eq. \ref{eq:dif1} will go to 
the constant $C$. This behavior, which is observed in Fig. 1, should be expected for any reasonable definition of 
$\sigma$ and $g(\sigma)$ and $s_{e}$ approximation. 

In order to avoid this pathology we could for example replace $\sigma$ as the preferred length scale with 
$l\propto 1/\rho\sigma^2$, the Boltzmann mean-free path. The new relationship is then:
\begin{eqnarray}
D^*=\frac{D}{D_B g_c(\phi)}=\exp{(C_0 s_{e}/k_B)}
\label{eq:dif2}
\end{eqnarray}
where we introduced a different constant $C_0 $. The test of this suggested dependence is shown in Fig. 2. 
The hard sphere results are very well represented by the new equation with $C_0=0.80$, while for the argon 
results a better fit is $C_0=0.83$. It may be interesting to test the validity of Eq. \ref{eq:dif2} for other systems as 
well, e.g. liquid metals \cite{has,dd02}. 
We also show for comparison the predictions of the effective 
diameter Enskog theory. The disagreement with the simulation results is similar 
with the one observed for the hard sphere fluid \cite{ew91}, and it is even bigger for MET (not shown).

The diffusion theory of Cohen and Turnbull \cite{ct} builds upon the physical concept of ``free-volume'' 
$v_f$ available for a molecule, originally introduced by Van der Waals to account for the effect 
of short range repulsive forces between molecules. The diffusion coefficient is written as:
\begin{eqnarray}
D=g a(v^*) v_T\exp{(-\gamma v^*/v_f)}
\end{eqnarray}
\begin{figure}
\includegraphics[width=8.0cm]{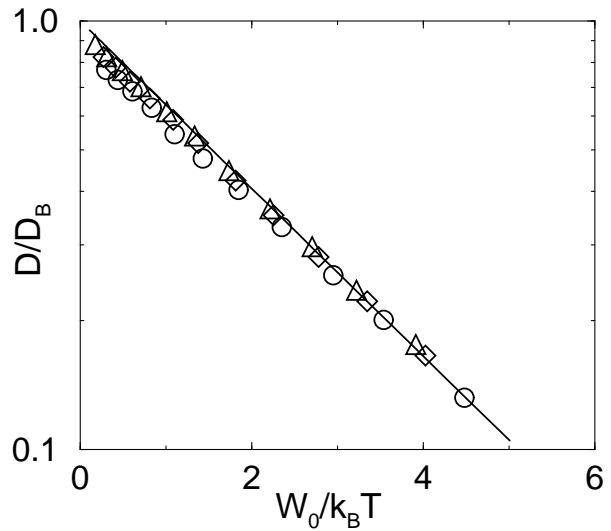}
\caption{Scaled diffusion coefficient of argon as a function of void producing work $W_0$ (see text - Eq. \ref{eq:dif4}). 
Symbols (same as Fig. 1): 
simulation results. Solid line: Eq. \ref{eq:dif4} with $\zeta=0.45$.}
\end{figure}
where $g$ is a geometric factor, $a(v^*)$ is roughly the diameter of the neighbor-induced ``cage'' inhabited by a 
molecule, $v_T$ is the thermal velocity and $v^*$ is 
essentially proportional with the molecular volume $v_0=\pi\sigma^3/6$ and can be identified with it 
with a suitable rescaling of the constant $\gamma$. This theory has been recently recast in a more transparent form 
as an Arrhenius theory of activation \cite{re02}. The transformation is done by 
recognizing first that in the limit of large free volumes relative to the molecular volume, $v_f\gg v_0$, the Boltzmann 
result for the diffusion coefficient should be recovered. Second, the free volume $v_f$ is expressed in terms of 
an effective pressure $p_r$ that includes only the repulsive (excluded volume) contributions, in the spirit of 
the Van der Waals theory: $p_r v_f = k_B T$. The proposed relation for $D$ is:
\begin{eqnarray}
\frac{D}{D_B}=\exp{(-\zeta W/k_BT)}
\label{eq:dif3}
\end{eqnarray}
where $\zeta$ is a constant and $W=p_r v_0$ is interpreted as the work necessary to create a void of volume $v_0$ in 
the liquid under the effective pressure $p_r$, to be occupied by the diffusing molecules. The ambiguity in defining 
this pressure is solved by recasting the usual pressure equation for a liquid into a generalized Van der Waals 
form \cite{re02}. There remains the task of defining a suitable hard core diameter $\sigma$, which can be avoided 
for the scaling factor by using $D_0$ instead of $D_B$, but it is required for $v_0$. Eq. \ref{eq:dif3} has therefore 
been applied only to systems modelled by interactions that explicitly include a hard core, e.g. hard sphere 
with an attractive 
square well \cite{re02}. We note that in fact Eq. \ref{eq:dif3} is unambiguously defined for a hard sphere fluid:
\begin{eqnarray}
&&\frac{D}{D_B}=\exp{(-\zeta W_0/k_BT)}
\label{eq:dif4}
\end{eqnarray}
where $W_0/k_B T=pv_0/k_BT=\phi(1+\phi+\phi^2-\phi^3)/(1-\phi)^3$ and we used the Carnahan-Starling equation for 
the pressure.
This can then be easily applied to typical Van der Waals-like potentials, 
\begin{figure}
\includegraphics[width=8.0cm]{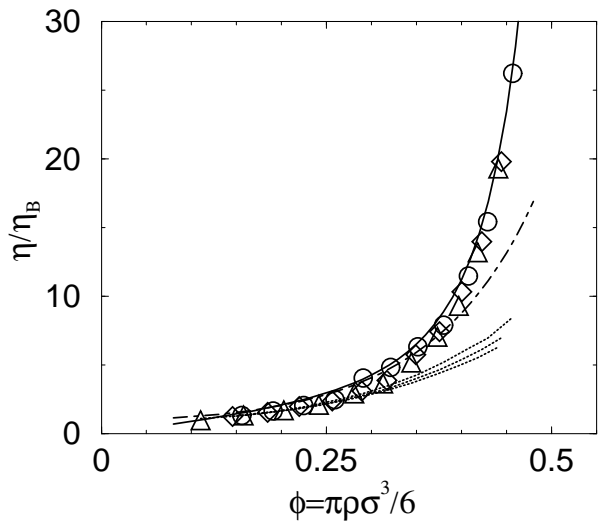}
\caption{Scaled viscosity of argon as a function of effective packing fraction (see text). Symbols (same as Fig. 1): 
simulation results. Solid line: Einstein relation with ``slip'' boundary condition ($c=2\pi$ - see text). Dot-dashed line: 
(effective diameter) Enskog theory. Dotted lines: modified Enskog theory (MET) - $T=298,1000,3000$K, top to bottom.}
\end{figure}
e.g. that of Eq. \ref{eq:exp6}, by using the hard 
sphere effective diameter given by Eq. \ref{eq:vart}. The scaled argon simulation results 
are shown in Fig. 3 as a function of $W_0/k_B T$. The data is well 
fitted by $\zeta=0.45$; a better representation is obtained with an additional prefactor, 
i.e. $A_0\exp{(-\zeta W_0/k_BT)}$, for 
both argon and the hard sphere fluid, albeit with different $\zeta's$. Nevertheless, as seen for example in Fig. 2, 
the mapping of the argon 
diffusion constant into that of the hard sphere fluid is more accurate than the Enskog theory except in a very narrow 
domain of intermediate densities.

We now turn to the discussion of the collective transport coefficients shear viscosity $\eta$ and thermal 
conductivity $\lambda$ using the same effective diameter approach. 
The Boltzmann-scaled simulation results for the argon viscosity, $\eta/\eta_B$, are shown in Fig. 4 as a function of 
the effective 
packing fraction $\phi$. Similarly with diffusion, single-variable scaling appears to hold rather well and the behavior of the 
viscosity largely mirrors that of the hard sphere fluid (not shown). Given that the Enskog theory 
strongly underestimates the hard sphere values at 
\begin{figure}
\includegraphics[width=8.0cm]{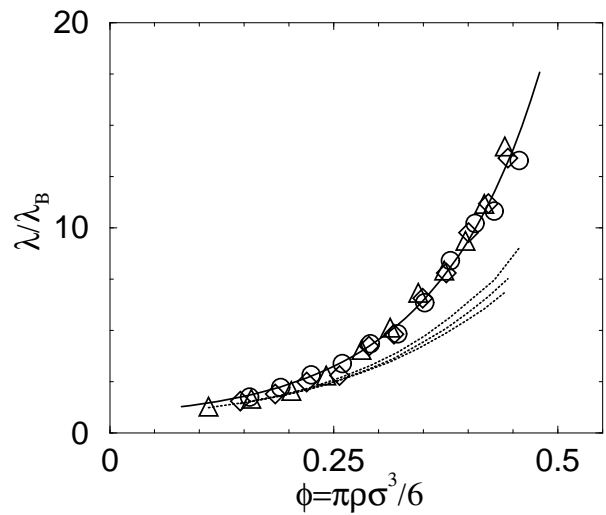}
\caption{Scaled thermal conductivity of argon as a function of effective packing fraction (see text). Symbols (same as Fig. 1): 
simulation results. Solid line: (effective diameter) Enskog theory. Dotted lines: modified Enskog theory (MET) - 
$T=298,1000,3000$K, top to bottom.}
\end{figure}
high densities \cite{agw,sh03}, it is not surprising 
that both adaptations of the theory, effective diameter Enskog and modified Enskog (MET), 
fail to capture the steep rise of $\eta$ as the system moves closer to freezing. 
Nevertheless, there are significant differences between the two approaches. The modified Enskog theory appears 
to work well at lower densities, as observed in \cite{hmc}, and slightly better than the effective diameter version. 
This however reverses quickly as the density increases, with the effective diameter method emerging as a much better 
estimator than MET at high densities. Although it may be interesting to pinpoint the origin of this different 
behavior, which also occurs for the thermal conductivity, this is difficult due to the convoluted nature of MET. 
It suffices perhaps to remark that a large part of the difference between the two procedures at high densities 
is due to a smaller $y$ (see Eqs. \ref{eq:enskog} and below) in the 
MET approach, which is proportional with the Enskog theory collision frequency.
The rapid increase of $\eta$ at large packings may be better reproduced by assuming that the diffusion constant is inversely 
proportional with the shear viscosity, i.e. the Einstein relation, $D=k_B T/c\eta \sigma$ \cite{agw}. 
We find that the 'slip' boundary condition $c=2\pi$ provides a reasonably good match to the $\eta$ dependence on 
$\phi$ in the dense region \cite{erd}, in agreement with \cite{agw}. 

Among the transport properties of the hard sphere system the thermal conductivity is most accurately predicted by the 
Enskog results up to the liquid-solid transition \cite{agw,sh03}. It is therefore important to 
assess if the success of 
the theory can also be transferred to fluids well described by Van der Waals-type potentials, e.g. argon. 
As shown in Fig. 5 the effective hard sphere 
diameter allows again a very good single-variable representation of the thermal conductivity for all 
temperatures studied. Moreover, the use of this 
diameter in the Enskog relation is successful in modeling the simulation results over the entire 
range of densities simulated. The modified Enskog theory (MET) on the other hand yields increasing discrepancies 
as the density rises and it is therefore not appropriate at high pressures. 

Finally, it would be desirable to test the above methods for the calculation of transport coefficients against
real dense fluids experimental data. Unfortunately, these are somewhat scarce. For example, only thermal 
transport measurements have been performed up to GPa pressures \cite{tn80,ngt}, and recently 
extended to tens of GPa just for the case of oxygen \cite{asb}. We limit ourselves therefore to the 
available high pressure thermal conductivity data and consider here argon ($Ar$), neon ($Ne$), nitrogen($N_2$) 
and oxygen ($O_2$). The first two of these are monoatomic fluids naturally modeled by isotropic potentials. 
The last ones are small diatomics, but a spherical interaction approximation turns out to be 
very successful in predicting the thermodynamics of these molecular systems in 
\begin{figure}
\includegraphics[width=8.0cm]{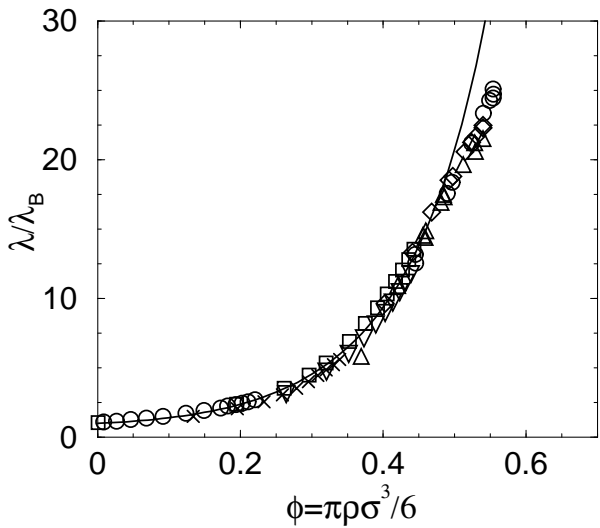}
\caption{Scaled experimental thermal conductivity as a function of effective packing fraction: triangles down - argon at 
$T=298$K from Ref. \cite{ngt}; crosses - neon at $T=298$K from Ref. \cite{ngt}; squares - nitrogen at $T=298$K 
from Ref. \cite{tn80}; circles - oxygen at $T=298$K from Ref. \cite{asb} (high pressures) and Ref. \cite{hmr82} (low pressures); 
diamonds - oxygen at $T=473$K from Ref. \cite{asb}; triangles up - oxygen at $T=573$K from Ref. \cite{asb}. 
Solid line: (effective diameter) Enskog theory.}
\end{figure}
a wide domain of temperatures 
and pressures \cite{rr80,sjh,br00}. All are well described by Buckingham {\it exponential-6} potentials with $\alpha=13.2$; 
the other parameters are: ($\epsilon^{Ar}/k_B=122K$, $r_{0}^{Ar}=3.85$\AA) \cite{rmbx} - also used in simulations, 
($\epsilon^{Ne}/k_B=42K$, $r_{0}^{Ne}=3.18$\AA) \cite{vsyr}, ($\epsilon^{N_2}/k_B=101.9K$, $r_{0}^{N_2}=4.09$\AA) \cite{rr80}, 
($\epsilon^{O_2}/k_B=125K$, $r_{0}^{O_2}=3.86$\AA) \cite{vtr}. The experimental results that we use for comparison 
are the ones of \cite{tn80,ngt,asb} and also \cite{hmr82}. The calculation of the effective diameters is 
done as before with the use of Eq. \ref{eq:vart} and results are shown in Fig. 6.

The success of single-variable scaling for both monoatomic and small diatomic molecules through the use of the 
effective hard sphere diameter is remarkable. The resulting master curve is also in very good agreement with 
the Enskog prediction in a large domain of packing fractions, which appears to roughly coincide with the 
equilibrium hard sphere fluid region that extends up to $\phi\simeq 0.494$. 
While a comparison between systems described by different types of interactions or at least Buckingham potentials 
with different $\alpha$'s would be a more stringent test of the existence 
of an universal curve for the scaled thermal conductivity, the agreement with the Enskog theory lends very 
good support to this idea for fluids that are well modeled by classical Van der Waals-like potentials. 

The disagreement at the largest $\phi$'s is rather interesting, particularly because the Enskog theory 
is known to slightly {\it underestimate} the thermal conductivity of the hard 
sphere system in the dense regime \cite{agw,sh03}. The corresponding experimental data 
have been recently obtained for dense oxygen \cite{asb}. Since the oxygen molecule is in fact anisotropic 
the observed discrepancy, where the effective diameter Enskog prediction significantly {\it overestimates} 
the experimental values, 
could be reasonably attributed to a breakdown of the spherical potential approximation at high densities. 
The fact that most calculated effective packing fractions lie in the metastable region of the hard sphere liquid, 
which is unexpected for an equilibrium fluid if it is fully modeled by Van der Waals-type interactions,  
also seems to support this idea \cite{note}. This breakdown however appears to be rather subtle because the 
thermodynamics 
based on Eq. \ref{eq:vart} and the Buckingham {\it exponential-6} potential still reproduces very well, 
within approximately $2\%$, all oxygen densities measured in the experiments.
 
Molecular dynamics calculations of the transport 
properties of hard ellipsoids, which should be a better approximation for the $O_2$ molecule at high densities, 
indicate that if the system is dense even a small molecular anisotropy {\it decreases} the thermal conductivity 
compared to the hard sphere system \cite{btae}. This can be understood intuitively as follows: for very dense systems the 
collisional contribution to thermal conduction is dominant \cite{eucken}, but the 
energy transfer in collisions is less ``efficient'' for hard anisotropic bodies than for 
isotropic ones. This ``efficiency'' is 
even further reduced for molecules such as $O_2$ that also posses vibrational degrees of freedom. 
For example, the energy exchange between translations and vibrations can involve exceedingly 
long relaxation times compared to those typical for translations alone \cite{cgp}. 
Such effects severely limit the usefulness of the hard sphere system as a reference for the 
description of transport properties of dense systems with 
multiple - translational, rotational, vibrational - degrees of freedom, even when that may still 
be appropriate for thermodynamics. This would also suggest that in this case, in addition to thermodynamics, 
structural properties may become increasingly important in determining dynamical behavior at 
high densities. With respect to the Enskog approach, the assumption of a single relaxation time appears 
already to be its major drawback for such systems, even when the anisotropy, for example, is partly 
accounted for \cite{btae}.

The preceding analysis indicates that the thermal conductivity of nitrogen ($N_2$), whose 
molecular size is comparable to that of $O_2$, will likely exhibit a similar behavior at high 
pressures when described in terms of an 
effective diameter, while even larger deviations should be expected for more 
anisotropic molecules, e.g. $CO_2$. Moreover, the above discussion should also apply to the viscosity 
{\it mutatis mutandis} \cite{btae}. High pressure experimental data on the transport properties of these or 
similar molecular systems, although difficult to obtain \cite{asb}, would help in understanding 
the interplay between thermodynamic and structural properties on the one hand, and transport behavior on 
the other, for very dense fluids. 

I would like to thank E. Abramson for kindly providing the data published in \cite{asb}. 
This work was performed under the auspices of the U. S. Department of Energy by 
University of California Lawrence Livermore National Laboratory under Contract 
No. W-7405-Eng-48.


\begin{thebibliography}{99}
\bibitem{erd}To this end we observe that the Einstein relation can be written as $\eta/\eta_B=7.873\phi/(D/D_B)$ 
and use the single variable fit for the $D/D_B$ argon simulation results given by 
$A_0\exp{(-\zeta W_0/k_BT)}$ with $\zeta=0.43$ and $A_0=0.92$; the other $D/D_B$ fits yield qualitatively similar 
predictions for $\eta/\eta_B$.
\bibitem{note}It is however not surprising that such large equivalent packing fractions still correspond to 
fluid oxygen in thermodynamic equilibrium. This is due to the fact that the real fluid has degrees of 
freedom which are unaccounted for in the spherical approximation, i.e. rotational and vibrational, 
resulting in a larger entropy than its spherically modeled counterpart.
\bibitem{eucken}We note that for this reason correcting the kinetic contribution to the thermal 
conductivity to account for the internal degrees of freedom as proposed by Eucken for polyatomic 
gases \cite{fk} is largely irrelevant at high densities; 
it would lead in fact to an extremely small increase in the overall Enskog theory prediction.
\end{thebibliography}
\end{document}